\documentclass[smallextended]{svjour3}
\usepackage{graphicx} 
\usepackage{amsfonts}
\usepackage{amssymb}
\newcommand{\eqref}[1]{\mbox{(\ref{#1})}} 
\begin {document}

\title{Averaging in LRS class II spacetimes}

\author{Petr Ka\v{s}par\textsuperscript{\textdagger}	\and
        Otakar Sv\'{\i}tek\textsuperscript{\S}
}

\institute{Institute of Theoretical Physics\\
Faculty of Mathematics and Physics, Charles University in Prague\\
V Hole\v{s}ovi\v{c}k\'{a}ch 2, 180 00 Praha 8, Czech Republic\\ 
              \email{\textsuperscript{\textdagger}petrkaspar@atlas.cz, \textsuperscript{\S}ota@matfyz.cz}           
}

\date{Received: date / Accepted: date}

\maketitle

\begin{abstract}
We generalize Buchert's averaged equations [Gen. Rel. Grav. 32, 105 (2000); Gen. Rel. Grav. 33, 1381 (2001)] to LRS class II dust model in the sense that all Einstein equations are averaged, not only the trace part. We derive the relevant averaged equations and we investigate backreaction on expansion and shear scalars in an approximate LTB model. Finally we propose a way to close the system of averaged equations.
\end{abstract}

\section {Introduction}

Our universe is considered to be homogeneous and isotropic on the large scale leading to the FLRW model. However, if we move to smaller scales, we can observe a strongly inhomogeneous distribution of structures. If we want to deal with inhomogeneity rigorously and at the same time keep a consistent connection with the FLRW geometry we may consider an averaging formalism to smooth out the metric tensor and at the same time average Einstein equations as well. The problem is that Einstein equations are nonlinear and if we average them straightforwardly we do not obtain an averaged metric tensor as a solution of averaged equations. Instead, we should consider an additional term --- the so-called correlation term, which can change the evolution of a smooth metric tensor and lead to the so-called backreaction. This term arises due to the nonlinearity of Einstein equations. It does not need to satisfy the usual energy conditions so it can possibly act as dark energy. 

While building a rigorous averaging scheme we face the problem that the average value of a tensor field is not well defined. There are several different approaches to define averages of tensors. One of the most promising ones is the scheme by Zalaletdinov~\cite{Zal1},~\cite{Zal2} where not only Einstein equations but also Cartan structure equations (and their integrability conditions) are averaged. A theorem about isometric embedding of a 2-sphere into Euclidian space is applied in the averaging method developed by Korzy\'{n}ski~\cite{Korz}. In~\cite{Brannlund} Weitzenb\"{o}ck connection for parallel transport is used to define the average value of a tensor field.

One of the most popular approaches to averaging is the one investigated by Buchert~\cite{Buchert1},~\cite{Buchert2}, where only the scalar part of Einstein equations is averaged. Wiltshire used this approach to give an alternative explanation of cosmic acceleration \cite{Wiltshire}. This theory was also applied to the cosmological perturbation theory \cite{Li1}, \cite{Li2}, \cite{Behrend}, \cite{Clarkson}. For observational issues see e.g. \cite{Larena}. In this paper, we will generalize Buchert's equations to the locally rotationally symmetric (LRS) class II dust family of spacetimes. The LRS family was classified in \cite{dust}, \cite{Stewart} and recently in \cite{LRS}. It contains e.g. LRS Bianchi cosmologies, Kantowski-Sachs model or LTB model and its generalizations. We will use the fact that this family is described by scalars to average the complete set of Einstein equations, including constraints. Although the averaged constraints are shown to be preserved during evolution the averaged system of equations is not closed and additional information has to be supplemented. 

In the past there have been many attempts to apply Buchert's approach to the LTB model. Papers \cite{Sussman} and \cite{Chuang} are comprehensive studies of Buchert's formalism applied to generic LTB models. For a treatment of Buchert's equations\ inside LTB spacetime see e.g. \cite{Paranjape1}, \cite{Rasanen1}, \cite{Bolejko2}, \cite{Sussman2}, \cite{Sussman3}, \cite{Mattsson1}, \cite{Mattsson2} and \cite{Mattsson3}. For an application of Buchert's formalism to the structure formation see e.g. \cite{Rasanen2}, \cite{Rasanen3} and \cite{Paranjape2}. 

Naturally, one can study inhomogeneities perturbatively on a homogeneous background and many important results are based on this approach. However, we should be cautious about relying solely on a linear perturbative analysis when dealing with a nonlinear theory. The effects of the correlation term indicate what kind of effects one might be missing when using a simple approach. In this sense, rigorous averaging of exact inhomogeneous spacetimes leading to standard cosmological models provides a possibility to qualitatively estimate these effects.

\vspace*{\baselineskip}
The paper is organized as follows. In Section 2 we review the LRS family and its characterizations, then we briefly mention Buchert's equations. Next, we average equations describing dust LRS class II family. After a short review of LTB metric in Section 5 we investigate the backreaction in the so-called onion model. We proceed by attempting to close the averaged equations and we finish with conclusion.

\section {LRS family}
Locally rotationally symmetric (LRS) dust spacetimes are defined by the following features \cite{dust}: In an open neighborhood of each point $p$, there exists a nondiscrete subgroup of the Lorentz group which leaves invariant the Riemann tensor and its covariant derivatives up to the third order. Therefore, in LRS spacetimes there exists a preferred direction $e^\mu$ (the axis of symmetry) at every point. The subgroup can be one or three-dimensional. In the latter case, we can rotate the axis of symmetry and spacetimes are everywhere isotropic - these are the FLRW models.

We will use the covariant 1+3 splitting of spacetime with the timelike vector field $u^\mu$ normalized by the condition $u_\rho u^\rho = -1$ and the projection tensor $h_{\mu \nu}= g_{\mu \nu} + u_\mu u_\nu$. In this section we will follow the article of van Elst and Ellis \cite{LRS}.

The preferred spacelike vector field $e^\mu$ satisfies the following conditions:
\begin{equation}
e_\rho u^\rho = 0, \,\,\,\,\,\, e_\rho e^\rho=1.
\end{equation} 
\indent Because of the defining property of the LRS spacetime, all covariantly defined spacelike vectors orthogonal to $u^\mu$ (acceleration $\dot u^\mu$, vorticity $\omega^\mu$, projected gradient of density $h_{\,\,\mu }^\sigma  \nabla _\sigma \rho $, pressure $h_{\,\,\mu }^\sigma  \nabla _\sigma  p$ and expansion  $h_{\,\,\mu }^\sigma  \nabla _\sigma  \theta$) must be proportional to $e^\mu$ - if this condition does not hold, spacelike vectors will not be invariant under the rotation about $e^\mu$.
\begin{equation}
\dot u^\mu   = \dot ue^\mu, \,\,\,\,\,\,\,\,\,\   \omega^\mu   =\omega e^\mu,
\end{equation}

\begin{equation}
 h_{\,\,\mu }^\sigma  \nabla _\sigma  \rho  = \rho 'e_\mu, \,\,\,\,\,\   h_{\,\,\mu }^\sigma  \nabla _\sigma  p  = p 'e_\mu, \,\,\,\,\,\  h_{\,\,\mu }^\sigma  \nabla _\sigma  \theta  = \theta 'e_\mu.
\end{equation}
Dots denote covariant derivative along the flow vector $u^\mu$ and primes denote covariant derivative along the vector $e^\mu$. We define the magnitude of the spatial rotation $k$ and the magnitude of the spatial divergence $a$ by
\begin{equation}
k: = \left| {\eta ^{\alpha \beta \gamma \delta } \left( {\nabla _\beta  e_\gamma  } \right)u_\delta  } \right|,
\end{equation}

\begin{equation}
a: = h_{\,\,\beta }^\alpha  \left( {\nabla _\alpha  e^\beta  } \right),
\end{equation}
where $\eta ^{\alpha \beta \gamma \delta }$ is the totally antisymmetric Levi-Civita pseudotensor ($\eta_{0123}=-\sqrt{-g}$). A similar rule works also for the spacelike tracefree symmetric tensors orthogonal to $u^\mu$.  Following \cite{LRS} we introduce a tensor field $e_{\mu \nu}$ defined by $e^\mu$
\begin{equation}
e_{\mu \nu } : = {1 \over 2}\left( {3e_\mu  e_\nu   - h_{\mu \nu } } \right).
\end{equation}
Then we have the relations for the shear tensor and the electric and magnetic parts of the Weyl tensor
\begin{equation}
\sigma _{\mu \nu }  = {2 \over {\sqrt 3 }}\sigma e_{\mu \nu }, \,\,\,\,\,\,\,\, E _{\mu \nu }  = {2 \over {\sqrt 3 }}E e_{\mu \nu }, \,\,\,\,\,\,\,\, 
H _{\mu \nu }  = {2 \over {\sqrt 3 }}H e_{\mu \nu }. 
\end{equation}
Here we can see that the LRS spacetimes are characterized only by a finite set of scalar functions.

For simplicity we will restrict our attention to the LRS class II dust models defined by the relation $k=\omega=0$. One can also show that the magnetic part of the Weyl tensor is equal to zero, $H=0$. This family of spacetimes includes the LTB metric and its generalizations based on foliation by spacelike 2-surfaces with negative or zero curvature scalar. The relevant evolution equations are
\begin{eqnarray}
\label{prvni}
\dot \theta  &=&  - {1 \over 3}\theta ^2  - 2\sigma ^2  - 4\pi \rho,  \\
\label{druha}
\dot \sigma  &=&  - {1 \over {\sqrt 3 }}\sigma ^2  - {2 \over 3}\theta \sigma  - E, \\
\dot E &=&  - 4\pi \rho \sigma  + \sqrt 3 E\sigma  - \theta E,\\
\dot \rho  &=&  - \rho \theta, \\
\label{pata}
\dot a &=&  - {1 \over 3}a\theta  + {1 \over {\sqrt 3 }}a\sigma,
\end{eqnarray}

and the constraints
\begin{eqnarray}
\label{sconstraint}
\sigma ' &=& {1 \over {\sqrt 3 }}\theta ' - {2 \over 3}a\sigma ,\\
\label{econstraint}
E' &=&  - {3 \over 2}aE + {{4\pi } \over {\sqrt 3 }}\rho ',\\
\label{posledni}
a' &=& {2 \over 9}\theta ^2  + {2 \over {3\sqrt 3 }}\theta \sigma  - {4 \over 3}\sigma ^2  - {2 \over {\sqrt 3 }}E - {1 \over 2}a^2  - {{16\pi } \over 3}\rho.
\end{eqnarray}
If we take the time derivative of the constraints, we can prove that they do not change with time.

\section {Buchert's equations}
In this section we will review an averaging method developed by Buchert \cite{Buchert1}. We will consider only the dust case - for generalization to perfect fluid see \cite{Buchert2}. This approach uses 1+3 splitting of spacetime, which is well defined by irrotational dust 4-velocity. However, averaging is well defined only for scalars, therefore only scalar part of Einstein equations is averaged. Given a scalar field $A$, the average value over three-dimensional spacelike domain ${\cal D}$ is defined by
\begin{equation}
\label{stred}
\left\langle A \right\rangle _{\cal D}  =  {1 \over {V_{\cal D} }}\int\limits_{\cal D} {d^3 XJA = } {1 \over {V_{\cal D} }}\int\limits_{\cal D} {d^3 X\sqrt {\det g_{ij} } A},
\end{equation}
where $J:=\sqrt{det g_{ij}}$, $g_{ij}$ is the metric of the spacelike hypersurface, $X^i$ are the comoving coordinates and $V_{\cal D}$ is the proper volume of the three-dimensional domain ${\cal D}$. From this definition we can see that time derivative and averaging do not commute. We have a commutation relation 
\begin{eqnarray}
\label{kom}
\left\langle A \right\rangle _{\cal D}^ \cdot   &=& {d \over {dt}}\left( {{1 \over {V_{\cal D} }}\int\limits_{\cal D} {d^3 XJA} } \right) =  - {{\dot V_{\cal D}} \over {V_{\cal D} }}\left\langle A \right\rangle _{\cal D}  + {1 \over {V_{\cal D} }}\int\limits_{\cal D} {d^3 X\left( {\dot JA + J\dot A} \right)} \nonumber\\
 &=& - \left\langle \theta  \right\rangle _{\cal D} \left\langle A \right\rangle _{\cal D}  + \left\langle {A\theta } \right\rangle _{\cal D}  + \left\langle {\dot A} \right\rangle _{\cal D},
\end{eqnarray}
where the expansion rate $\theta$ is related to the velocity of the fluid $u^\mu$ according to the definition $\theta=u^\mu_{\,\,\,;\mu}$. Next, in analogy with the FLRW spacetime we introduce the dimensionless scale factor $a_{\cal D}$ and the effective Hubble parameter $H_{\cal D}$
\begin{equation}
{a_{\cal D}}=\left(\frac {V_{\cal D}}{V_{{\cal D}_i}}\right) ^{\frac{1}{3}},
\end{equation}
\begin{equation}
\left\langle \theta \right\rangle_{\cal D}=\frac {\dot{V_{\cal D}}}{V_{\cal D}}=3\frac {\dot{a_{\cal D}}}{a_{\cal D}}=:3H_{\cal D}.
\end{equation}
\indent $V_{{\cal D}_i}$ is the volume of the initial domain which is geodetically evolved into $V_{\cal D}$. Now we have a formalism for averaging scalars. To obtain a scalar equation from Einstein equations, we have to contract them with available tensors
 - i.e. $g^{\mu \nu}, u^\mu$ and $\nabla^\mu$. After contracting we get the Raychaudhuri equation, the Hamiltonian constraint and the continuity equation. Now we perform averaging and use the commutation rule \eqref{kom}
\begin{equation}
3 \frac{\ddot{a_{\cal D}}}{a_{\cal D}}+4\pi G \left\langle \rho \right\rangle _{\cal D} = {\cal Q}_{\cal D},
\label{raych}
\end{equation}
\begin{equation}
\left(\frac{\dot{a}_{\cal D}}{a_{\cal D}} \right)^2-\frac{8 \pi G}{3} \left\langle \rho \right\rangle _{\cal D}+\frac{\left\langle {\cal R} \right\rangle _{\cal D}}{6}=-\frac{{\cal Q}_{\cal D}}{6},
\label{hamilton}
\end{equation}
\begin{equation}
\partial_t \left\langle \rho \right\rangle _{\cal D}+3\frac {\dot{a_{\cal D}}}{a_{\cal D}}\left\langle \rho \right\rangle _{\cal D}=0.
\label{kont}
\end{equation}
${\left\langle {\cal R} \right\rangle _{\cal D}}$ denotes the average value of the spatial Ricci scalar, ${\left\langle {\rho} \right\rangle _{\cal D}}$ means the average density of the fluid and ${\cal Q}_{\cal D}$ that shows possible backreaction (due to inhomogeneity and anisotropy) is defined by
\begin{equation}\label{backreactionQ}
{\cal Q}_{\cal D} := {2 \over 3}\left\langle {\left( {\theta  - \left\langle \theta  \right\rangle _{\cal D} } \right)^2 } \right\rangle _{\cal D}  - 2\left\langle {\sigma ^2 } \right\rangle _{\cal D}.
\end{equation}
The scalar $\sigma^2={1 \over 2}\sigma_{\mu \nu} \sigma^{\mu \nu} $ is constructed from the shear tensor.

\section {Averaging LRS class II dust spacetime}
Now, we will generalize the above approach to LRS class II dust solutions. Originally, Buchert considered spacetimes with a dust \cite{Buchert1} or a perfect fluid \cite{Buchert2} source. He did not assume any symmetries or simplifications and his equations can be applied to a large class of metrics. Here we will restrict to spacetimes with the special LRS symmetry. For this family we will generalize Buchert's equations in the sense that all Einstein equations are averaged consistently.

Given a preferred spacelike direction $e_\mu$, all the equations describing the LRS metric are scalar. It means that we can perform averaging (which is covariantly defined for scalars). We will define averaging over the spacelike domain ${\cal D}$ according to \eqref{stred}. In order to obtain averaged equations we need to derive commutation relations for the time and spatial derivatives (with respect to the preferred direction). For the LRS class II spacetime it is possible to express derivative along the preferred direction $\vec e$ by the formula $\vec e= \sqrt{g_{11}} \partial_1 $, where $g_{11}$ is the metric function of a particular solution inside the LRS class II spacetime. The most studied solution inside this class is the LTB model, for which the square root of the metric function reads $\sqrt{g_{11}}= \frac{R'}{\sqrt{1+2B}}$. We will show the basic facts about the LTB metric in the next section. The commutation rule between coordinate derivative and averaging reads
\begin{equation}
\partial_1 \left\langle A \right\rangle - \left\langle \partial_1 A \right\rangle =  \left\langle A  \frac{\partial_1J}{J} \right\rangle - \left\langle A \right\rangle \left\langle \frac{\partial_1J}{J}  \right\rangle.
\end{equation}

For simplicity we will restrict to the class II LRS spacetime with the condition $p=0 \Leftrightarrow \dot{\rho}= - \rho \theta$ (dust models) which includes LTB spacetimes and their generalizations. For simpler notation we shall omit symbol $\cal D$ at the averaging bracket in the rest of the paper (however we retain the symbols not defined as direct averages - $a_{\cal D}, {\cal Q}_{\cal D}$). If we average the equations ~\eqref{prvni} -~\eqref{posledni} we obtain
\begin{eqnarray}
\label{Buchert}
\left\langle \theta  \right\rangle ^ \cdot   &=&  - {1 \over 3}\left\langle \theta  \right\rangle ^2  - 4\pi \left\langle \rho  \right\rangle  + \underline{{2 \over 3}\left( {\left\langle {\theta ^2 } \right\rangle  - \left\langle \theta  \right\rangle ^2 } \right) - 2\left\langle {\sigma ^2 } \right\rangle }\ , \\
\label{shear}
 \left\langle \sigma  \right\rangle ^ \cdot   &=&  - {1 \over {\sqrt 3 }}\left\langle \sigma  \right\rangle ^2  - {2 \over 3}\left\langle \theta  \right\rangle \left\langle \sigma  \right\rangle  - \left\langle E \right\rangle  + \underline {{1 \over {\sqrt 3 }}\left( {\left\langle \sigma  \right\rangle ^2  - \left\langle {\sigma ^2 } \right\rangle } \right)} \nonumber\\
& & + \underline{ {1 \over 3}\left( {\left\langle {\theta \sigma } \right\rangle  - \left\langle \theta  \right\rangle \left\langle \sigma  \right\rangle } \right)}\ , \\
\label{E}
 \left\langle E \right\rangle ^ \cdot   &=&  - 4\pi \left\langle \rho  \right\rangle \left\langle \sigma  \right\rangle  + \sqrt 3 \left\langle E \right\rangle \left\langle \sigma  \right\rangle  - \left\langle \theta  \right\rangle \left\langle E \right\rangle \nonumber\\
 & & \underline{ - 4\pi \left( {\left\langle {\rho \sigma } \right\rangle  - \left\langle \rho  \right\rangle \left\langle \sigma  \right\rangle } \right) + \sqrt 3 \left( {\left\langle {E\sigma } \right\rangle  - \left\langle E \right\rangle \left\langle \sigma  \right\rangle } \right)}\ , \\
\label{hmota}
\left\langle \rho  \right\rangle ^ \cdot   &=&  - \left\langle \rho  \right\rangle \left\langle \theta  \right\rangle\ , \\
\left\langle a \right\rangle ^ \cdot   &=&  - {1 \over 3}\left\langle a \right\rangle \left\langle \theta  \right\rangle  + {1 \over {\sqrt 3 }}\left\langle a \right\rangle \left\langle \sigma  \right\rangle  + \underline{ {2 \over 3}\left( {\left\langle {a\theta } \right\rangle  - \left\langle a \right\rangle \left\langle \theta  \right\rangle } \right) } \nonumber\\
& & + \underline{ {1 \over {\sqrt 3 }}\left( {\left\langle {a\sigma } \right\rangle  - \left\langle a \right\rangle \left\langle \sigma  \right\rangle } \right) }\ , \\
\label{constrain1}
\left\langle \sigma \right\rangle ' &=&  \frac{1}{\sqrt{3}} \left\langle \theta \right\rangle' - \frac{2}{3} \left\langle a \right\rangle \left\langle \sigma \right\rangle + \underline{\sqrt{g_{11}} \left( \left\langle \sigma \frac{\partial_1J}{J} \right\rangle - \left\langle \sigma \right\rangle \left\langle \frac{\partial_1J}{J} \right\rangle \right)} \\
&-& \underline{\sqrt{\frac{g_{11}}{3}} \left( \left\langle \theta \frac{\partial_1J}{J} \right\rangle + \left\langle \theta \right\rangle \left\langle \frac{\partial_1J}{J} \right\rangle \right) -  \frac{2}{3} \left( \sqrt{g_{11}} \left\langle \frac{1}{\sqrt{g_{11}}} a \sigma  \right\rangle - \left\langle a \right\rangle \left\langle \sigma \right\rangle \right) }\ , \nonumber\\
\label{constrain2}
\left\langle E \right\rangle ' &=&  - \frac{3}{2} \left\langle a \right\rangle \left\langle E \right\rangle + \frac{4 \pi}{\sqrt{3}} \left\langle \rho \right\rangle ' + \underline{ \sqrt{g_{11}} \left( \left\langle E \frac{\partial_1J}{J} \right\rangle  - \left\langle E \right\rangle  \left\langle \frac{\partial_1J}{J} \right\rangle  \right) } \\
&-& \underline{ 4\pi \sqrt{\frac{g_{11}}{3}} \left( \left\langle \rho \frac{\partial_1J}{J} \right\rangle - \left\langle \rho \right\rangle \left\langle \frac{\partial_1J}{J} \right\rangle  \right) - \frac{3}{2} \left( \sqrt{g_{11}} \left\langle \frac{1}{\sqrt{g_{11}}} a E  \right\rangle - \left\langle a \right\rangle \left\langle E \right\rangle \right) }, \nonumber\\
\label{constrain3}
\left\langle a \right\rangle ' &=& \frac{2}{9}\left\langle \theta \right\rangle^2 + \frac{2}{3\sqrt{3}} \left\langle \theta \right\rangle \left\langle \sigma \right\rangle - \frac{4}{3} \left\langle \sigma \right\rangle^2 - \frac{2}{\sqrt{3}} \left\langle E \right\rangle - \frac{1}{2} \left\langle a \right\rangle^2 - \frac{16 \pi}{3} \left\langle \rho \right\rangle \\
&+& \underline{\sqrt{g_{11}} \left( \left\langle a \frac{\partial_1J}{J} \right\rangle  - \left\langle a \right\rangle  \left\langle \frac{\partial_1J}{J} \right\rangle  \right) + \frac{2}{9} \left( \sqrt{g_{11}} \left\langle \frac{1}{\sqrt{g_{11}}} \theta^2  \right\rangle - \left\langle \theta \right\rangle^2 \right) } \nonumber\\
&+& \underline{\frac{2}{3\sqrt{3}} \left( \sqrt{g_{11}} \left\langle \frac{1}{\sqrt{g_{11}}} \theta \sigma  \right\rangle - \left\langle \theta \right\rangle \left\langle \sigma \right\rangle \right) -  \frac{4}{3} \left( \sqrt{g_{11}} \left\langle \frac{1}{\sqrt{g_{11}}} \sigma^2  \right\rangle - \left\langle \sigma \right\rangle^2 \right)  } \nonumber\\
&-& \underline{\frac{2}{\sqrt{3}} \left( \sqrt{g_{11}} \left\langle \frac{1}{\sqrt{g_{11}}} E  \right\rangle - \left\langle E \right\rangle \right) - \frac{1}{2} \left( \sqrt{g_{11}} \left\langle \frac{1}{\sqrt{g_{11}}} a^2  \right\rangle - \left\langle a \right\rangle^2 \right)} \nonumber\\
&-& \underline{\frac{16 \pi}{3} \left( \sqrt{g_{11}} \left\langle \frac{1}{\sqrt{g_{11}}} \rho  \right\rangle - \left\langle \rho \right\rangle \right)}\ . \nonumber   
\end{eqnarray} 
The underlined parts of the equations denote additional terms due to averaging. We can recognize the well-known Buchert's equation~\eqref{Buchert} with the kinematical backreaction term and the mass conservation equation~\eqref{hmota}. 

If we want to restrict the above equations to the LTB model (the line element is given in the next section) we can substitute the following expressions for the magnitude of the spatial divergence $a$, its average value $\left\langle a \right\rangle_{\cal D}$ and the square root of the spatial part of the metric $J=\sqrt{det g_{ij}}$
\begin{equation}
a=h^\nu_\mu e^\mu_{\,\,\,\,;\nu}= \frac{2\sqrt{1+2B}}{R}, \,\,\,\, \left\langle a \right\rangle_{\cal D} = \frac{4 \pi R^2}{V_{\cal D}}, \,\,\,\, J= \frac{R'R^2 \sin{\theta}}{\sqrt{1+2B}}.  
\end{equation}

 One can show that the averaged constraint equations~\eqref{constrain1} -~\eqref{constrain3} are preserved in time. The key role in the calculation is played by the equation \cite{LRS}
 \begin{equation}
 \left( {f'} \right)^ \cdot   = \left( {\dot f} \right)^\prime   - {2 \over {\sqrt 3 }}\sigma f' - {1 \over 3}\theta f', 
 \label{dotprime}
 \end{equation}
 and its averaged version. Now, we can take time derivative of the constraints~\eqref{constrain1} -~\eqref{constrain3}. Using commutation rules and equations ~\eqref{prvni} -~\eqref{posledni}, a slow but straightforward computation will show that the constraints do not evolve in time. The explicit computation for equation (\ref{constrain1}) is shown in the Appendix.

All of Einstein equations are averaged now. It means that we can investigate not only backreaction on the expansion rate but also on shear scalar or electric part of the Weyl scalar. The problem is that the equations are not closed. We need additional relations to close the system because for example $\left\langle \theta \right\rangle$ is independent of $\left\langle \theta^2 \right\rangle$. In the next chapters we will give some suggestions for closing the system of equations. 

So far we have not seen an analog of the Hamiltonian constraint. For example Sussman in \cite{Sussman} used three-dimensional curvature $\cal R$ instead of the function $a$. The relation between three-dimensional curvature $\cal R$ and the magnitude of the spatial divergence $a$ reads \cite{LRS}
\begin{equation}
 {\cal R} = - \left( 2 a' + \frac{2}{3} a^2 - 2 K \right),
\end{equation}
where $K$ is the Gaussian curvature of the 2-D spacelike group orbits orthogonal to $e^\mu$ and $u^\mu$. The Hamiltonian constraint has the form 
\begin{equation}
 {\cal R} = 16 \pi \rho - \frac{2}{3} \theta^{2} + 2 \sigma^{2}.  
\end{equation}
The averaged Hamiltonian constraint is Buchert's equation \eqref{hamilton}
\begin{eqnarray}
\left\langle  {\cal R} \right\rangle  &=& 16 \pi \left\langle  \rho \right\rangle - \frac{2}{3} \left\langle \theta \right\rangle^2 + 2\left\langle  \sigma \right\rangle^{2} \nonumber\\
& & \underline{ - \frac{2}{3} \left( \left\langle \theta^2 \right\rangle - \left\langle \theta \right\rangle^2 \right) + 2 \left( \left\langle \sigma^2 \right\rangle - \left\langle \sigma \right\rangle^2 \right)}.
\end{eqnarray}

\section {LTB metric}
The most important representative of the dust LRS class II family is the LTB spacetime. In this section we will briefly review its properties. 

The Lema\^{i}tre-Tolman-Bondi (LTB) metric~\cite{Lemaitre},~\cite{Tolman},~\cite{Bondi} is a spherically symmetric exact solution of Einstein equations. It corresponds to an inhomogeneous dust with the stress energy tensor 
\begin{equation}
T_{\mu \nu}=\rho u_\mu u_\nu, 
\end{equation}
where $u_\mu$ is 4-velocity of the dust with density $\rho$. For a recent review of LTB metric, see e.g.~\cite{Bolejko},~\cite{Hellaby}. The line element reads
\begin{equation}
ds^2=-dt^2+\frac{(R')^2}{1+2B(r)}dr^2+R^2(t,r)[d \theta^2 +\mbox{sin}^2( \theta) d \phi^2],
\end{equation}
where $B(r)$ is an arbitrary function and the prime denotes partial derivative with respect to $r$. Function $R(t,r)$ is a solution of Einstein equation
\begin{equation}
\label{LTB}
R_{,t}^2=2B+ \frac{2M}{R} + \frac{\Lambda}{3} R^2.
\end{equation}
$M=M(r)$ is another arbitrary function of integration. The energy density $\rho$ is determined by the equation
\begin{equation}
4 \pi \rho = \frac{M'}{R'R^2}.
\label{LTBrho}
\end{equation}
The function $B(r)$ is related to the quasi-local spatial curvature \cite{Sussman} and $M(r)$ is the gravitational mass contained within a comoving spherical shell at a given $r$. Equation~\eqref{LTB} can be integrated to yield
\begin{equation}
\int\limits_0^R {{{d\tilde R} \over {\sqrt {2B + {{2M} \over {\tilde R}} + {1 \over 3}\Lambda \tilde R^2 } }}}  = t - t_B (r).
\end{equation}
Here $t_B(r)$ is a third free function of $r$ (called the bang time function). In the LTB model, in general, the Big Bang is not simultaneous as in the FRW case, but it depends on the radial coordinate $r$. The given formulas are invariant under a transformation $\tilde{r}=g(r)$. We can use this freedom to choose one of the functions $B(r), M(r)$ and $t_B(r)$. For $\Lambda=0$ the above equation can be solved explicitly. The evolution can be elliptic ($B<0$), parabolic ($B=0$) or hyperbolic ($B>0$).

\section {Backreaction inside the LTB onion model}
As an example of backreaction computation we consider an approximate LTB model (the so-called onion model) investigated in~\cite{Biswas} by Biswas, Mansouri and Notari, who computed corrections to the luminosity distance--redshift relation. 
\begin{figure}
  \centering
  \includegraphics[width=7cm]{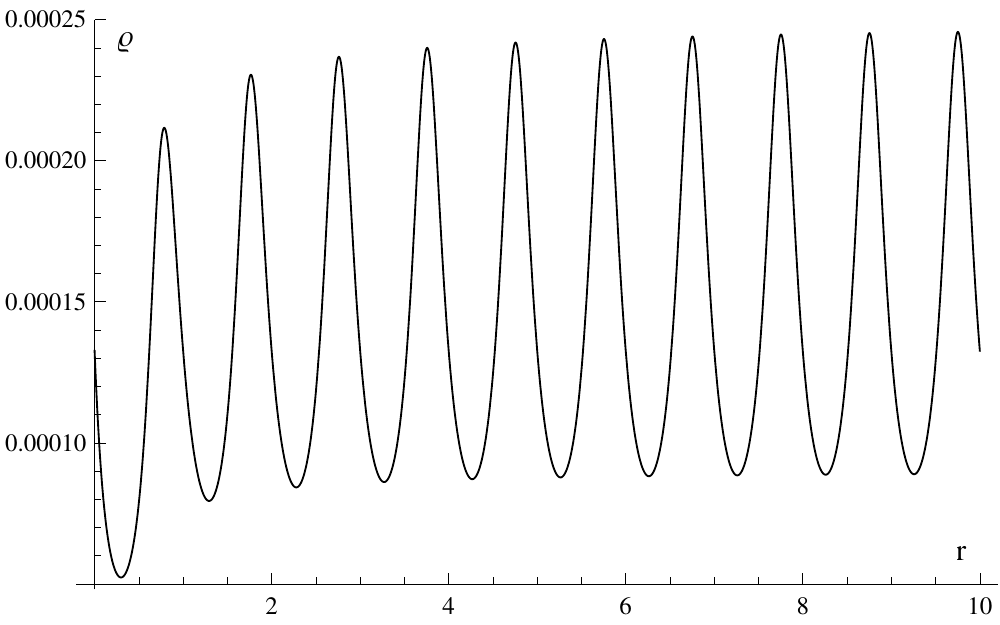}
  \caption{Density profile at the time $t=20$.}
  \label{fig:rho}
\end{figure}
It represents a spacetime with radial shells of overdense and underdense regions. The function $B(r)$ is nonzero ($B(r)>0$), so the evolution of the LTB model is hyperbolic. The metric function $R(t,r)$ reads
\begin{equation}
\label{onion}
R(t,r):=\left(\frac{6}{\pi}\right)^{1/3}t^{2/3}r \left( 1+ \left( \frac{81}{4000 \pi ^2} \right)^{1/3} \left( \frac{1}{2 \pi} \right)  t^{2/3} \frac{1}{r} \sin^2{\pi r} \right).
\end{equation}
The function $B(r)$ is given as follows
\begin{equation}
B(r)= \frac{r}{2 \pi} \sin^2{\pi r}.
\end{equation}
The density profile at the time $t=20$ can be seen in Figure~\ref{fig:rho}. The density is computed using the formula \eqref{LTBrho}. The coordinates were chosen so that the function $M(r)$ is given by $M(r)=4/3 \pi r^3$. 

\begin{figure}
  \centering
	\includegraphics[width=7cm]{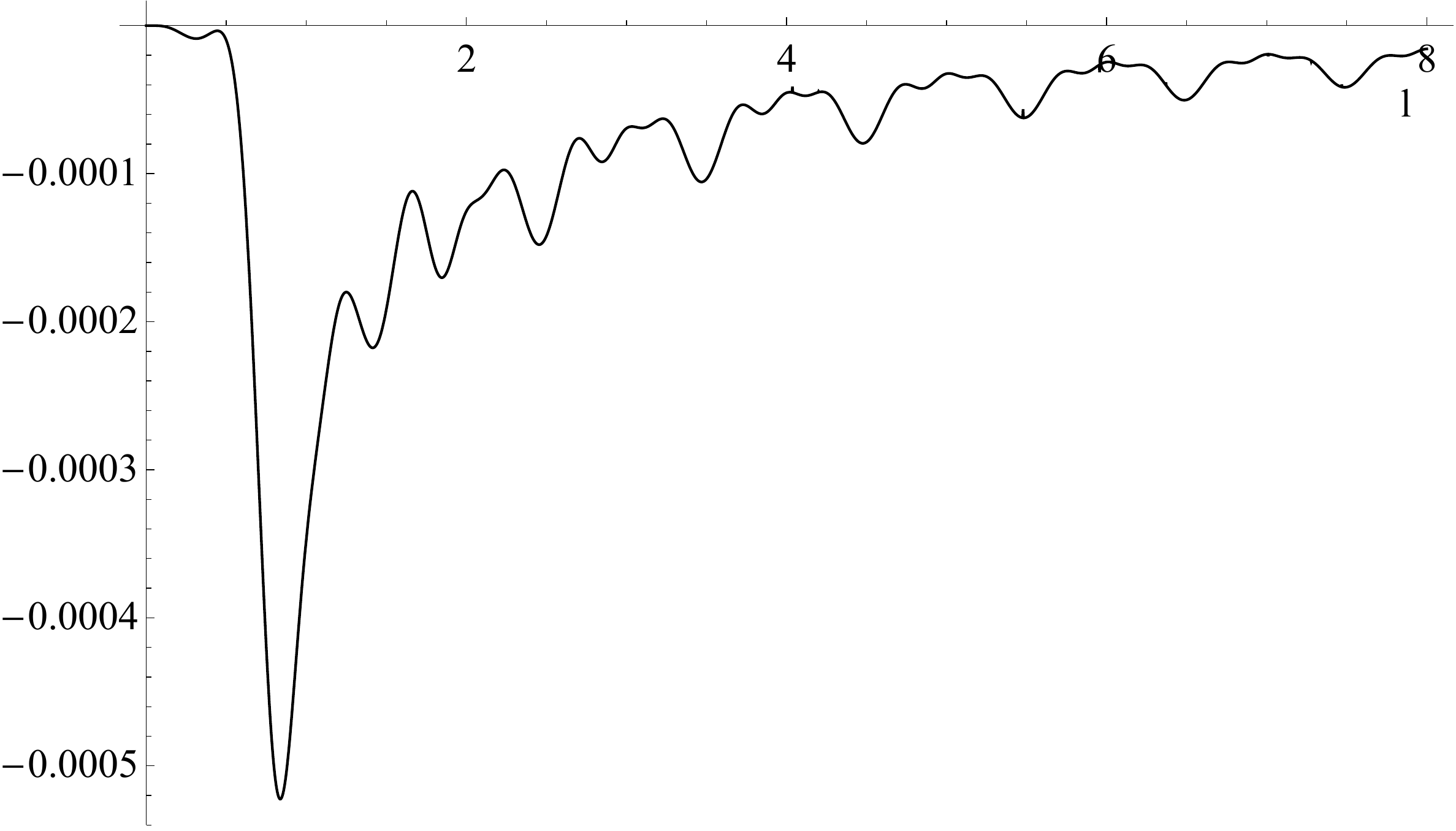}
  \caption{Backreaction term ${2 \over 3}\left( {\left\langle {\theta ^2 } \right\rangle  - \left\langle \theta  \right\rangle ^2 } \right) - 2\left\langle {\sigma ^2 } \right\rangle$ in the evolution equation for expansion depending on the averaging scale $l$ and normalized by $\left\langle \theta \right\rangle ^ \cdot$.} 
  \label{fig:exp}
\end{figure}

\begin{figure}
  \centering
	\includegraphics[width=7cm]{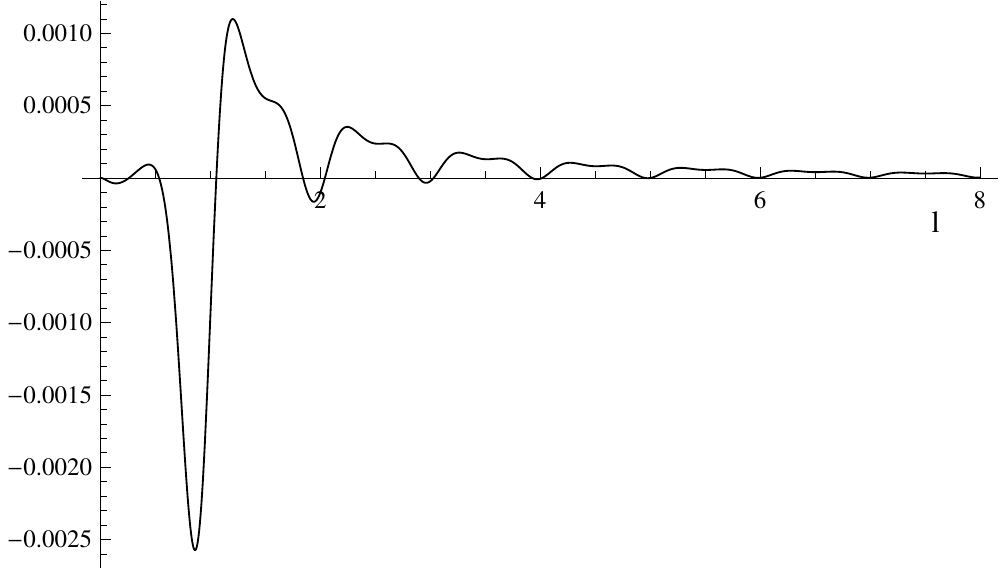}
  \caption{Backreaction term ${1 \over {\sqrt 3 }}\left( {\left\langle \sigma  \right\rangle ^2  - \left\langle {\sigma ^2 } \right\rangle } \right) +  {1 \over 3}\left( {\left\langle {\theta \sigma } \right\rangle  - \left\langle \theta  \right\rangle \left\langle \sigma  \right\rangle } \right)$ in the evolution equation for shear depending on the averaging scale $l$ and normalized by $\left\langle \sigma \right\rangle ^ \cdot$.} 
  \label{fig:shear}
\end{figure}

First, we investigate the backreaction term in Buchert's equation \eqref{Buchert}. We numerically integrate the underlined part of equation \eqref{Buchert} depending on the averaging scale $l$.  As one can see from Figure~\ref{fig:exp}, the backreaction normalized by $\left\langle \theta \right\rangle ^ \cdot$ is negative. It has a peak for the averaging scale $l\approx0.8\, .$ The value of backreaction normalized by $\left\langle \theta \right\rangle ^ \cdot$ is of the order of $10^{-4}$. The backreaction term (without normalization) is positive and it leads to an increase of expansion. 
We can investigate also the backreaction terms in other equations which do not appear in the Buchert framework and which can supplement his equations. For example here we will show the result for backreaction in the averaged evolution equation for shear \eqref{shear} (specifically the whole underlined part of the equation is considered). All results depend on the averaging scale $l$. As we can see from Figure~\ref{fig:shear} - for small scales, the contribution of all backreaction terms in the evolution equation for shear normalized by $\left\langle \sigma \right\rangle ^ \cdot$ is negative with a peak around $l\approx 0.9\, .$ For larger scales the contribution is smaller and positive. The turning point is for $l\approx 1.2\, .$ To be more precise, there exist regions where the backreaction changes the sign twice for a very small increase of $l$. If we compare the backreaction with the time derivative of the shear scalar, we can see that their ratio is of the order of $10^{-4} - 10^{-3}$. It means that the backreaction plays a more important role in the averaged equation for shear than in the averaged equation for expansion. 

Note that we investigated only an approximate LTB model. Due to nonlinearity it is not clear if the backreaction behavior shown above will be similar for exact solutions (even when they are close to the onion model in some specific sense). We used this non-exact model because it has suitable properties for investigation of averaging and backreaction. 

\section {Averaged LRS dust class II equations}
One of the most important equations in cosmology is the evolution equation for the expansion scalar. In the averaged equation \eqref{Buchert} we have independent variables $\left\langle {\theta ^2 } \right\rangle$ and $\left\langle {\sigma ^2 } \right\rangle$. To obtain evolution equation for $\left\langle {\theta ^2 } \right\rangle$ we multiply \eqref{prvni} by $2 \theta$. Then we perform averaging and we obtain the equation
\begin{eqnarray}
\label{theta2}
 \left\langle \theta^2 \right\rangle ^ \cdot &=& -\frac{2}{3}\left\langle \theta \right\rangle^3 - 4 \left\langle \theta \right\rangle \left\langle \sigma \right\rangle^2 - 8 \pi \left\langle \rho \right\rangle \left\langle \theta \right\rangle + \underline{\frac{1}{3} \left( \left\langle \theta^3 \right\rangle - \left\langle \theta \right\rangle^3 \right)} \\
	& &\underline{ + \left( \left\langle \theta \right\rangle^3 - \left\langle \theta \right\rangle \left\langle \theta^2 \right\rangle \right) - 4 \left( \left\langle \theta \sigma^2 \right\rangle - \left\langle \theta \right\rangle \left\langle \sigma \right\rangle^2 \right) - 4 \pi \left( \left\langle \rho \theta \right\rangle - \left\langle \rho \right\rangle \left\langle \theta \right\rangle \right)    }. \nonumber
\end{eqnarray}
In a similar way we derive an evolution equation for $\left\langle {\sigma ^2 } \right\rangle$
\begin{eqnarray}
\label{sigma2}
 \left\langle \sigma^2 \right\rangle ^ \cdot &=& -\frac{2}{\sqrt{3}}\left\langle \sigma \right\rangle^3 - \frac{4}{3} \left\langle \theta \right\rangle \left\langle \sigma \right\rangle^2 - 2 \left\langle E \right\rangle \left\langle \sigma \right\rangle + \underline{\frac{1}{3} \left( \left\langle \theta \right\rangle \left\langle \sigma \right\rangle^2 - \left\langle \theta \sigma^2 \right\rangle \right)} \nonumber\\
 & & \underline{ - \frac{4}{3} \left( \left\langle \theta \right\rangle  \left\langle \sigma^2 \right\rangle - \left\langle \theta \right\rangle  \left\langle \sigma \right\rangle^2  \right) - \frac{2}{\sqrt{3}} \left( \left\langle \sigma^3 \right\rangle - \left\langle \sigma \right\rangle^3 \right) } \nonumber\\
& & - \underline{ 2 \left( \left\langle E \sigma \right\rangle - \left\langle E \right\rangle \left\langle \sigma \right\rangle \right) }.  
\end{eqnarray}
Now, we also need equations for e.g. $\left\langle {\theta ^3 } \right\rangle$,  $\left\langle {\theta \sigma ^2 } \right\rangle$ or $\left\langle {\sigma ^3 } \right\rangle$ (and of course an evolution equation for $\left\langle \rho \right\rangle$, $\left\langle E \right\rangle$ and $\left\langle \sigma \right\rangle$ given in Section 4). We could obtain these evolution equations by the same procedure. Thus we have an infinite number of equations for the correlation terms. Here we need to adopt an ansatz. For example we can consider a reasonable assumption that for a given order the correlation terms are negligibly small and we can truncate the hierarchy to obtain a finite set of equations. We can also assume that some terms are proportional to each other. In this approach the inhomogeneities are modeled by different relations for correlation functions.

The question is what kind of spacetime may correspond to the given set of averaged equations. We have started with scalar equations characterizing a LRS class II dust spacetime (containing the LTB metric and its generalizations). By averaging we can not leave this class, instead we may end up in a special subclass of LRS class II dust models. We performed averaging of the evolution equation for expansion \eqref{Buchert} and of the evolution equations for different products of expansion, shear, density and electric part of the Weyl tensor. From this construction we can see that the averaged equations contain an averaged LTB model, but generally not e.g. the homogeneous LRS Bianchi cosmologies. 

In the above described approach we have evolution equations for averages of different powers and products of the expansion, shear, density and electric part of the Weyl tensor. The problem is that if we derive an evolution equation for the averaged nonlinear terms, then more complicated terms appear in the relevant correlation terms (as is evident in equations (\ref{theta2}) and (\ref{sigma2})). If we want to close the system of equations we need to effectively eliminate these higher-order terms. In general, we may express the "unwanted terms" as a suitable function of the lower-order averaged terms whose evolution equation is known. This kind of ansatz does not need to make all higher-order correlation terms necessarily vanish, it only serves to close the system of averaged equations through the selected relations - these may be for example expressed as products of the averaged terms of lower order.

\section {Conclusion}
We generalized Buchert's equations for the LRS class II dust model. We used the property that this family is characterized only by scalars and we employed a similar technique for averaging. However, the system of averaged equations is not closed. Buchert considered the so-called scaling solutions \cite{Buchert3} to close the system of equations. In our work, we first investigated the influence of backreaction on the expansion and shear scalars for an approximate LTB model which describes fluctuating radial inhomogeneities. Then we proposed an infinite system of equations which supplement the averaged equations for expansion. In this approach inhomogeneities are modeled by the form of the correlation terms. Finally, we discussed how to close the system of averaged equations.

\section*{Acknowledgments}
We would like to thank R. Sussman for a useful discussion. We would also like to express our gratitude to the referee for correcting and improving our paper. P.K. was supported by grants GAUK 398911 and SVV-267301. O.S. acknowledges the support of grant GA\v{C}R 14-37086G.

\section*{Appendix}
\begin{appendix}

\setcounter{equation}{0}

In this appendix we will show the computations demonstrating that the averaged constraint equations ~\eqref{constrain1} -~\eqref{constrain3} are preserved in time. We start with an unaveraged constraint \eqref{sconstraint} and perform straightforward averaging without using the commutation rules which gives us
\begin{equation}\label{A1}
\left\langle \sigma ' \right\rangle = {1 \over {\sqrt 3 }} \left\langle \theta '  \right\rangle- {2 \over 3} \left\langle  a\sigma \right\rangle.
\end{equation}
Now we take time derivative of \eqref{A1} and use the commutation rule \eqref{kom}. We obtain the following expression
\begin{eqnarray}
\left\langle (\sigma')^ \cdot \right\rangle - \left\langle \theta \right\rangle \left\langle \sigma' \right\rangle + \left\langle \theta \sigma' \right\rangle &=& \frac{1}{\sqrt{3}}\left[ \left\langle (\theta')^ \cdot \right\rangle - \left\langle \theta  \right\rangle \left\langle \theta ' \right\rangle + \left\langle \theta \theta '  \right\rangle \right] \nonumber\\
 & & - \frac{2}{3} \left[ \left\langle (a \sigma)^\cdot \right\rangle - \left\langle \theta \right\rangle \left\langle a \sigma \right\rangle + \left\langle \theta a \sigma \right\rangle \right].
\end{eqnarray}
Now we need to commute prime and dot derivatives. This is done applying the commutation rule \eqref{dotprime}
\begin{eqnarray}
& &\left\langle (\dot{\sigma})' \right\rangle - \frac{2}{\sqrt{3}} \left\langle \sigma \sigma' \right\rangle - \frac{1}{3}  \left\langle \theta \sigma' \right\rangle - \left\langle \theta \right\rangle \left\langle \sigma' \right\rangle + \left\langle \theta \sigma' \right\rangle = \\
& &\frac{1}{\sqrt{3}}\left[ \left\langle (\dot{\theta})' \right\rangle - \frac{2}{\sqrt{3}}  \left\langle \sigma \theta' \right\rangle - \frac{1}{3}  \left\langle \theta \theta' \right\rangle - \left\langle \theta  \right\rangle \left\langle \theta ' \right\rangle + \left\langle \theta \theta '  \right\rangle \right]  - \frac{2}{3} \left[ \left\langle (a \sigma)^\cdot \right\rangle - \left\langle \theta \right\rangle \left\langle a \sigma \right\rangle + \left\langle \theta a \sigma \right\rangle \right]. \nonumber
\end{eqnarray}
Next, we apply the unaveraged evolution equations for $\theta, \sigma$ and $a$ (\eqref{prvni}, \eqref{druha} and \eqref{pata}) and we obtain the following expression:
\begin{eqnarray}\label{A4}
& & -\frac{2}{\sqrt{3}} \left\langle \sigma \sigma' \right\rangle - \frac{2}{3} \left\langle \theta' \sigma \right\rangle - \frac{2}{3} \left\langle \theta \sigma' \right\rangle - \left\langle E'\right\rangle  - \frac{2}{\sqrt{3}} \left\langle \sigma \sigma' \right\rangle - \frac{1}{3}  \left\langle \theta \sigma' \right\rangle - \left\langle \theta \right\rangle \left\langle \sigma' \right\rangle + \left\langle \theta \sigma' \right\rangle = \nonumber\\
& & \frac{1}{\sqrt{3}}\left[ - \frac{2}{3} \left\langle \theta \theta' \right\rangle - 4 \left\langle \sigma \sigma' \right\rangle - 4 \pi \left\langle \rho' \right\rangle  - \frac{2}{\sqrt{3}}  \left\langle \sigma \theta' \right\rangle - \frac{1}{3}  \left\langle \theta \theta' \right\rangle - \left\langle \theta  \right\rangle \left\langle \theta ' \right\rangle + \left\langle \theta \theta '  \right\rangle \right] \\
& & - \frac{2}{3} \left[ - \frac{1}{3} \left\langle a \theta \sigma \right\rangle + \frac{1}{\sqrt{3}} \left\langle a \sigma^2 \right\rangle - \frac{1}{\sqrt{3}} \left\langle a \sigma^2 \right\rangle - \frac{2}{3} \left\langle a \theta \sigma \right\rangle - \left\langle a E \right\rangle - \left\langle \theta \right\rangle \left\langle a \sigma \right\rangle + \left\langle \theta a \sigma \right\rangle \right]. \nonumber
\end{eqnarray}
From the above expression we can see that several terms cancel each other. Moreover, using constraint equations \eqref{sconstraint} and \eqref{econstraint} to further simplify the above equation \eqref{A4} it is straightforward to see that the left hand side is equal to the right hand side. This means that the constraint equation \eqref{constrain1} does not change in time. In the same way it can be shown that the constraint equations \eqref{constrain2} and \eqref{constrain3} are preserved in time, too. 
\end{appendix}

\def\bibname{Bibliography}

\end {document}